\documentclass[reprint,prd,onecolumn,notitlepage,nofootinbib,11pt]{revtex4-1}
\usepackage{amsmath,amssymb,graphicx,bm,lipsum,multirow}
\usepackage[colorlinks=true,citecolor=blue,linkcolor=blue,urlcolor=blue]{hyperref} 
\usepackage[font={small},flushleft,indent]{caption}
\usepackage{subcaption}
\usepackage{placeins}
\usepackage{overpic}

\begin{document}

\title{Detectability of secondary images from flares 
 near Sgr~A* with mock GRAVITY data}
\author{Fengting Xie}
\email{xiefengting@stu.cqu.edu.cn}
\author{Qing-Hua Zhu}
\email{zhuqh@cqu.edu.cn (corresponding author)}
\author{Xin Li}
\email{lixin1981@cqu.edu.cn (corresponding author)}
\affiliation{Department of Physics and Chongqing Key Laboratory for Strongly Coupled Physics, Chongqing University, Chongqing 401331, China}

\begin{abstract} 
The orbital motion of near-infrared flares reported by the GRAVITY collaboration encodes information about both the dynamics of accretion matter and the underlying spacetime geometry. The centroid track of these flares, which corresponds to the flux-weighted center of light, incorporates contributions from primary, secondary and higher-order images. Thus, it potentially indicates distinctive signatures of the spacetime geometry, even when these individual multiple images remain unresolved. 
In this study, we explore the detectability of the secondary images from flares orbiting Sgr~A* through mock data simulating future GRAVITY observations. Specifically, we compare the model in which the centroid coincides with the track of the primary images with another model in which the centroid incorporates flux-weighted contributions from both the primary and secondary images. Fitting these models to the mock data based on Bayesian framework, we quantify the conditions under which the signature of secondary images can be statistically distinguishable. We demonstrate that increasing the sample size by an order of magnitude alone could not yield strong evidence for distinguishing the secondary image. Robust detectability ($|\Delta\text{BIC}| >7.9$) is achieved when both with the improved sample size and astrometric uncertainties reduced to 40\% of current uncertainties of GRAVITY astrometric data. 
Unlike the primary image, which is dominated by accretion flow physics, the secondary images originate from gravitational lensing 
in the strong-field regime. Their detection is an essential first step toward probing higher-order images and the photon rings.
\end{abstract}

\maketitle

\section{Introduction}
Among currently accessible targets, Sagittarius A* (Sgr~A*) provides one of the promising laboratories for probing the strong-field regime of gravity. 
It is widely believed that a supermassive black hole (SMBH) resides at the center of our Galaxy \cite{Genzel:2010zy}, with a mass of $M=4.3\times10^6 ~M_{\odot}$ \cite{massofblackhole} and a distance of $D=8.178~\text{kpc}$ \cite{Genzel:2010zy}. 
As the nearest known SMBH, Sgr~A* has been the focus of continuous observations over the past decades \cite{genzel2021forty,Genzel:2010zy,Ghez:2003qj,milkywayimage}. 
In 2022, the Event Horizon Telescope (EHT) Collaboration released the first horizon-scale image of Sgr~A* \cite{milkywayimage,EHT:2022wok}, revealing its black hole shadow and providing support for the validity of general relativity (GR) at event-horizon scales \cite{Perlick:2021aok,Vagnozzi:2022moj,Zhang:2024jrw,Chen:2023wzv,Kuang:2024ugn,Li:2020drn,Chang:2021ngy,Zhu:2022shb,He:2024qka}. 
As a complement to test GR in the weak-field region far from the black hole \cite{Bertotti:2003,Confrontation:GRExperiment,2017:GReffectS2,abuter2018redshift,2019:redshifts0-2,GRAVITY:2020gka,Yao:2026pjz},  Sgr~A* provides a unique laboratory for studying accretion physics and testing Einstein's gravity in the strong-field regime \cite{narayan2023black,Levis:2023tpb,Galishnikova:2022mjg}.
In addition to horizon-scale imaging, Sgr~A* can also be probed through high-precision astrometric observations of infrared flares \cite{Abuter:2018uum,GRAVITY:2023data}. Notably, resolving the fine structure of black hole images offers a probe of spacetime geometry that is independent of accretion models \cite{KumarWalia:2024omf,Chen:2025jay,Capobianco:2025lxn,Hod:2026jtr}. This capability characterized by the photon rings, which are narrow, distinct features formed by photons executing multiple half-orbits, denoted $n$, around the black hole before reaching the observer \cite{Perlick:2004tq,Bozza:2010xqn,Gralla:2019xty,Salehi:2024cim,Tsupko:2025hhf}. 
The upcoming next-generation Event Horizon Telescope (ngEHT) \cite{doeleman2023reference,Deliyski:2024wmt} and 
the proposed Black Hole Explorer (BHEX) \cite{johnson2024black,Farah:2024mkq} mission may be capable of detecting the photon rings of Messier 87* (M87*) and Sgr~A*.
 
Although current observations are broadly consistent with the black hole interpretation, the detection of direct image of accretion disk alone seemingly does not uniquely confirm the existence of a conventional black hole in Einstein's gravity. Besides, the black hole found in modified gravity \cite{Wang:2025ihg,Lim:2025cne,Kobialko:2025jer,Eiroa:2025mws,Vagnozzi:2022moj,lupsasca2024beginner,Yin:2025rao}, various exotic compact objects have been proposed as black hole mimickers, including boson stars \cite{Vincent:2015xta,Olivares:2018abq}, gravastars \cite{Sakai:2014pga,PhysRevD.99.044027}, wormholes \cite{Shaikh:2018kfv,Kar:2024ctd,Macedo:2025ipc}, and fermionic dark-matter cores \cite{Pelle:2024eyt}. 
In particular, boson stars can produce direct image and even photon ring structures similar to those expected from Kerr black holes \cite{Vincent:2015xta}, while fermionic dark-matter cores might exhibit a central brightness depression surrounded by a ring-like feature lacking conventional photon rings \cite{Pelle:2024eyt}. 
This suggests that the direct image alone might not provide a definitive signature. Although photon rings may provide a more robust way to distinguish black holes from alternative objects \cite{Gralla:2019xty,Perlick:2021aok}, current observational precision prevent them from being resolved in the images \cite{EventHorizonTelescope:2019dse,milkywayimage,Johnson:2019ljv}. 
Recently, the detectability of secondary images, also referred to as the $n=1$ subring or leading subring, via unresolved observations remains an ongoing area of research, such as those based on image variability or correlation analysis  \cite{Chesler:2020gtw,Hadar:2020fda,Chen:2022kzv,Cardenas-Avendano:2024sgy,Wong:2024gph,Zhang:2025vyx,Zhu:2025jqh,Palumbo:2025tyu}. These approaches are expected to serve as an essential step prior to directly resolving photon rings.  

The GRAVITY instrument is a high-precision near-infrared interferometer, with the Galactic Center as one of its primary targets \cite{GRAVITYFirstlight}. It enables the centroid position of near-infrared flares near Sgr~A* to be tracked \cite{GRAVITYFirstlight,Abuter:2018uum,GRAVITY:2023data}. 
These flares are inferred to arise at distances of only a few Schwarzschild radii from the central object, making them potential probes of black hole spacetime in the strong-field regime \cite{GRAVITY:2023data}. Based on GRAVITY datasets, various kinematic scenarios for the flare motion have been investigated \cite{Jiang:2025huk,orbitalorpatternmotion,antonopoulou2024parameter,GRAVITY:2020lpa,Yfantis:2024eab,Wang:2026teu}. 
In this context, we explore the detectability of the secondary images in the near-infrared band with GRAVITY observations \cite{GRAVITY:2023data}. It could provide a complementary to studies in the millimeter band \cite{EventHorizonTelescope:2019dse,EHT:2022wok}, where analyses of image variability have not yet yielded observational evidence for the leading subring \cite{Cardenas-Avendano:2024sgy}. For the near-infrared band, the optical depth due to synchrotron self-absorption at higher frequencies is expected to be sufficiently low that it does not obscure the subring structure. In this sense, these flares therefore provide a promising opportunity to distinguish secondary or even higher-order images. 
While the GRAVITY instrument has demonstrated remarkable capabilities since its first light in 2017 \cite{GRAVITYFirstlight}, its current observational capacity remains limited by sample size and astrometric precision. This prevents the resolution of the secondary image, motivating our use of mock data to explore future detectability. In this study, we generate mock data characterized by increased sample sizes and enhanced astrometric precision, anchored to the reported GRAVITY flare event \cite{GRAVITY:2023data}. 
And we will quantify the conditions under which the signature of secondary images in Sgr~A* flares becomes statistically detectable.



The remainder of this paper is organized as follows. In Sec.~\ref{physical Setup and Ray Classification}, we present the physical background and the ray-tracing scheme, and define the observables, specifically the centroid tracks of flares. In Sec.~\ref{Data and Methodology}, we describe the procedure for generating mock data based on original GRAVITY datasets, and the implementation of the Bayesian inference for the flare motion. In Sec.~\ref{results}, we present our results, and in Sec.~\ref{Discussion and Conclusions}, we discuss and summarize our findings.

\section{Ray tracing method and centroid tracks} \label{physical Setup and Ray Classification}

In this study, we consider a spherically symmetric black hole located at the coordinate origin, whose spacetime metric can be written as
\begin{equation}
    \textrm{d}s^2=-f(r)\textrm{d}t^2+\frac{\textrm{d}r^2}{f(r)}+r^2(\textrm{d}\theta^2+\sin\theta^2\textrm{d}\phi^2)~.
\end{equation}
where $f(r)=1-2M/r$. Given that the current GRAVITY astrometric data provide limited constraint on the black hole spin, we adopted a non-rotating Schwarzschild black hole model in this study \cite{GRAVITY:2020gka,Yfantis:2024eab}.

Since the flare events occur at distances of only a few tens of Schwarzschild radii from the black hole, gravitational light bending cannot be neglected in the observations. We therefore employ the ray-tracing method developed in Ref.~\cite{hotspotmodel12}, which determines the trajectories of light rays by directly solving the null geodesics connecting the emission source and the observer. Compared with conventional backward ray-tracing algorithms, this approach is computationally more efficient for point-like or compact emission regions. 

In the strong-field region near a black hole, photons emitted from the same hotspot can reach a distant observer through multiple trajectories. These trajectories are classified by the image order $n$, which labels the half-orbit of the photon path around the black hole before reaching the observer.
\begin{figure}
    \centering
    \includegraphics[width=1\linewidth]{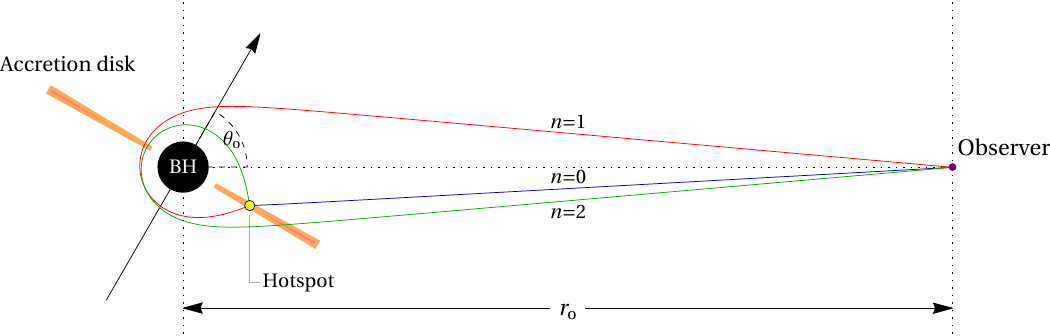}
    \caption{Schematic diagram of photon trajectories with different ray orders. Photons emitted from a hotspot on the accretion disk can reach the observer through multiple strongly lensed paths. The least-deflected path is labeled as $n=0$ and gives the primary image, while paths with additional winding around the black hole correspond to secondary ($n=1$) and tertiary ($n=2$) images. The observer is located at radius $r_o$ with inclination angle $\theta_o$.}
    \label{fig:rayorder}
\end{figure}
A schematic diagram of photon trajectories with different image orders is shown in Fig.~\ref{fig:rayorder}. The trajectory labeled by $n=0$ corresponds to the primary image.  It represents the least-deflected photon path connecting the source and the observer. The trajectory labeled by $n=1$ corresponds to the secondary image, also referred to as the leading subring \cite{Johnson:2019ljv,Palumbo:2025tyu}. Trajectories with larger integers of $n$ experience stronger gravitational bending. 

\begin{figure} 
    \centering
    \includegraphics[width=0.7\linewidth]{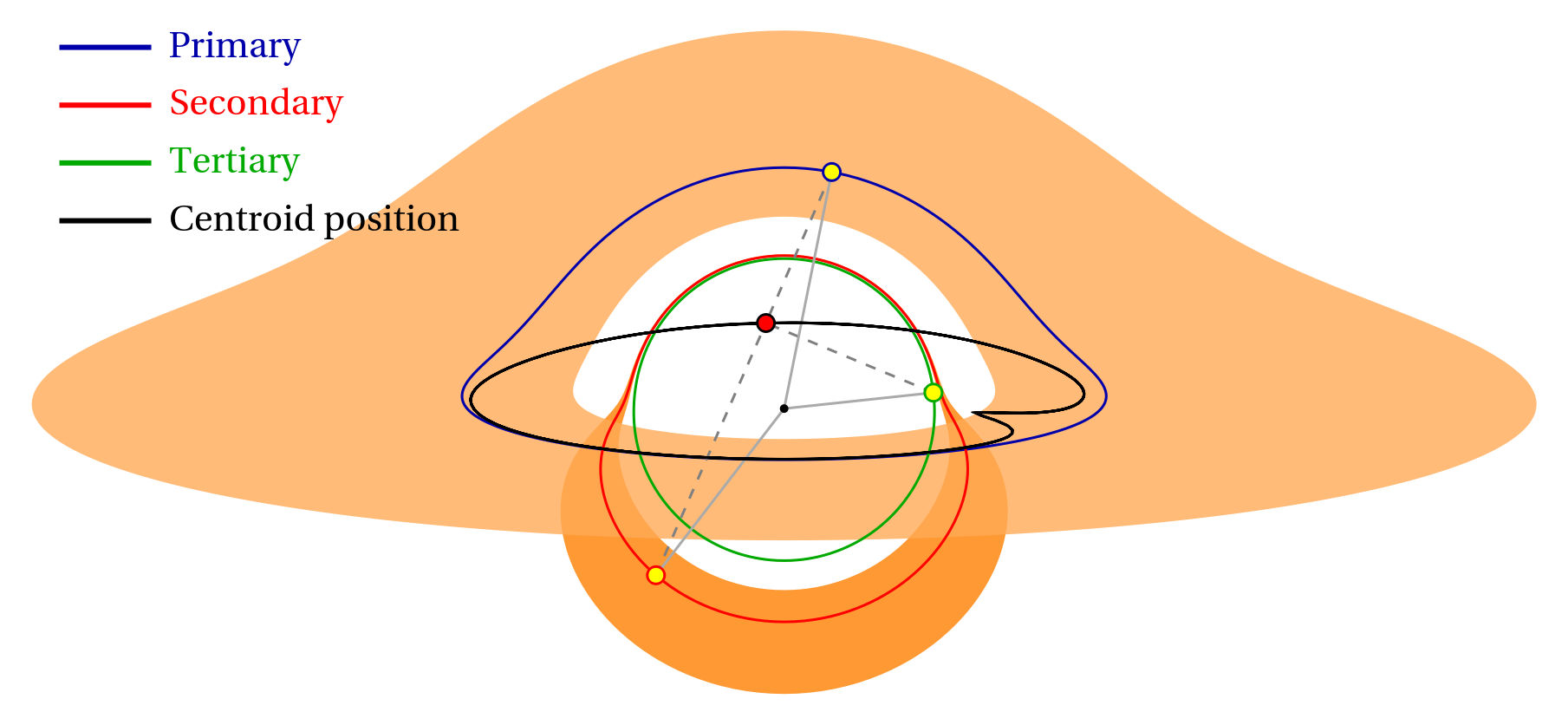}
    \caption{
Schematic diagram illustrating the centroid track. The yellow points represent the primary, secondary, and tertiary images. 
Yellow points denote the lensed images of flares confined to and corotating within the accretion disks. In this case, the images move along closed paths, shown as colored curves. The centroid position, denoted by the red point, is determined by the flux-weighted positions of the individual tracks of the primary, secondary, and higher-order images, given in Eq.~(\ref{3}). The centroid track, shown as the black curve, is distinguishable from the track of the primary images (blue curve). In this paper, we adopt the radiation model, assuming the hotspots distributed on the surface of accretion disks \cite{Bao:1992,1994ApJ...425...63B,hotspotmodel12}}
    \label{fig:ct}
\end{figure}

The GRAVITY instrument achieves microarcsecond astrometric precision, enabling highly accurate positional measurements near Sgr~A*. However, its angular resolution is limited to the milliarcsecond scale, meaning it cannot spatially resolve multiple distinct images at the microarcsecond scale \cite{GRAVITYFirstlight}.
In this context, one of the observables in the GRAVITY datasets is the centroid track of the Sgr~A* flares, which is defined as the flux-weighted average of the image positions. 
The observed flux $F^{(n)}_{\nu}$ for the $n$-th order image of a point source at near-infrared frequency band $\nu$ takes the form of
\begin{equation}
    F^{(n)}_\nu=\int_{\Omega_n} I_\text{obs}^{(n)}\mathrm{d}\Omega~,
\end{equation}
where the observed intensity takes the form of $I_\text{obs}=g^3 I_\text{em}$, $g$ is redshift factor,  $I_{\rm em}$ is the emitted specific intensity, and $\Omega$ denotes the solid angle on the observer's celestial sphere. 
Under the small-angle approximation, the celestial coordinates reduce to two-dimensional planar coordinates $\bm X =(X,Y)$.
At each observation time, the centroid position is formulated by \cite{hotspotmodel12,GRAVITY:2020lpa}
\begin{equation}
    \bm X^\text{centroid} \equiv \int    {\bm X} \bar{I}_\text{obs}(\bm X) \mathrm{d}^2{\bm X}  = \sum_{n=0}^{\infty} \bm X^{(n)} \bar{F}_\nu^{(n)}~. \label{3}
\end{equation}
Specifically, we have $\bm X^\text{centroid}=(X^\text{centroid},Y^\text{centroid})$ in the two-dimensional plane.
The second equality is obtained based on the point-like sources $\bar{I}_\text{obs}^{(n)} = \bar{F}_\nu^{(n)} \delta^2(\bm X - \bm X^{(n)}(t))$ \cite{Bao:1992,1994ApJ...425...63B,Bozza:2010xqn,Zhu:2025jqh,hotspotmodel12}, where $X^{(n)}(t)$ denotes the apparent track of the hotspot for given image order $n$, and the $\bar{F}_\nu^{(n)} [\equiv F^{(n)}_\nu/\sum_m F^{(m)}_\nu]$ is the normalized radiation flux. In this paper, we adopt the radiation model, assuming the hotspots distributed on the surface of accretion disks \cite{Bao:1992,1994ApJ...425...63B,hotspotmodel12}. In Fig.~\ref{fig:ct}, we illustrate the difference among the tracks of primary, secondary, tertiary images and the centroid tracks. Our results show that the centroid position differs from the trajectory of the primary images and can be modulated by the angular velocity of the accretion disk.
In principle, the centroid track should take account all the order of images.
For the first step, the purpose of this work is to test whether the secondary images ($n=1$) can be identified from centroid tracks in future GRAVITY datasets. Thus we focus on centroid tracks contributed by the primary image and the secondary image.



\section{Data and Methodology}\label{Data and Methodology}


Between 2018 to 2023, GRAVITY observed a total of eight flare events from the Galactic Center \cite{GRAVITY:2023data}. Among these events, two flare events have both polarization measurements and astrometric data. Of the remaining six flares, three have polarization measurements only, while the other three provide astrometric information only. 
All flares with polarization measurements exhibit a single clockwise loop, or a partial loop, in the $\text{Q-U}$ plane. For flares with astrometric data, the centroid exhibits clockwise motion in the plane of the sky with a period of approximately one hour, accompanied by full rotations of the polarization vector  at the same time. Given the high degree of similarity among different flare events in these observational properties, the astrometric data was averaged reasonably \cite{GRAVITY:2023data}. In this study, our mock data is generated based on the averaged astrometric data.

Physically, these flares are believed to originate from the dynamical processes of accreting material in the vicinity of the central black hole \cite{hotspotphysic1,hotspotphysic2}. Their phenomenological behavior can be described by a compact plasma blob orbiting near the innermost stable circular orbit (ISCO), commonly referred to as the hotspot model \cite{Bao:1992,Li:2014coa,hotspotmodel3,hotspotmodel4,hotspotmodel5,hotspotmodel6,hotspotmodel7,hotspotmodel8,hotspotmodel9,hotspotmodel10,hotspotmodel11,hotspotmodel12,hotspotmodel13,Zhao:2025ouq}. In this work, we adopt the hotspot model to phenomenologically describe the kinematics of the flares, in which the observed emission is assumed to arise from a localized bright spot corotating with the accretion disk \cite{Bao:1992,hotspotmodel12,Genzel:2010zy}. Previous studies have shown that smaller hotspot sizes generally lead to better fits to the data; therefore, consistent with our previous work, we consider a point-like hotspot model in this study \cite{GRAVITY:2020lpa,Xie:2025}.

Based on the results of our previous analysis, circular orbits within the equatorial plane provide statistically preferred fits to the GRAVITY astrometric flare data among planar geodesic models \cite{GRAVITY:2020lpa,Xie:2025}. Accordingly, in the following analysis, we restrict our orbital model to circular orbits. 

\subsection{Mock data generation}\label{mock data generation}

It is expected that increasing the number of observations and improving the measurement precision may enhance the ability to distinguish centroid tracks from different image orders. In this section, we thus construct mock astrometric datasets based on the best-fitting orbital parameters inferred from the GRAVITY observations to investigate how data quality affects the distinguishability of the primary and secondary images.

We adopt the best-fitting orbits inferred from the GRAVITY flare data as fiducial models, denoted as $(X^\text{fiducial}(t), Y^\text{fiducial}(t))$.  Within the time duration of the GRAVITY observations, we sample the centroid tracks $(X^\text{fiducial}(t_i), Y^\text{fiducial}(t_i))$ at discrete and equally spaced time series $t_i$, where $i=1,2,...,N$, and $N$ is desired sample size of the mock data. 
The uncertainties and the deviation from the fiducial values $X^\text{fiducial}(t_i)$ and $Y^\text{fiducial}(t_i)$ are constructed from the astrometric uncertainties reported in the GRAVITY flare observations, denoted as $\sigma^\text{obs}_{X,a}$ and $\sigma^\text{obs}_{Y,a}$ for $a=1,2,...,9$\footnote[1]{There are nine data points in the combined fit by GRAVITY collaboration \cite{GRAVITY:2023data}\label{footnote}}. 
First, the uncertainties in the mock data are described by the following probability distribution, 
\begin{align}
    \sigma_{X,i} \sim \mathcal{N}_{+}\left(\bar{\sigma}_X,s_X\right), \hspace{1cm}
    \sigma_{Y,i} \sim \mathcal{N}_{+}\left(\bar{\sigma}_Y,s_Y\right), \label{4}
\end{align}
where $\mathcal{N}_{+}(\mu,s)$ denotes a normal distribution with mean $\mu$ and standard deviation $s$, truncated to positive values. The center values in the distribution $\bar{\sigma}_X$ and $\bar{\sigma}_Y$ are given by $\bar{\sigma}_X = f\,{\rm Mean}(\sigma_{X,a}^\text{obs})$ and $\bar{\sigma}_Y = f\,{\rm Mean}(\sigma_{Y,a}^\text{obs})$, where we consider a scaling factor $f$ for the uncertainties of the mock data. Similarly, the standard deviation in the distributions ${s}_X$ and ${s}_Y$ are derived from $s_X = f\,{\rm Std}(\sigma_{X,a}^\text{obs})$ and $s_Y = f\,{\rm Std}(\sigma_{Y,a}^\text{obs})$. Secondly, the astrometric deviations relative to the fiducial values are modeled as independent Gaussian random variables, $\delta X_i \sim \mathcal{N}\left(0,\sigma_{X,i}\right)$ and $\delta Y_i \sim \mathcal{N}\left(0,\sigma_{Y,i}\right)$,  with standard deviations $\sigma_{X,i}$ and $\sigma_{Y,i}$ derived from Eqs.~(\ref{4}) via sampling. Based on the astrometric deviation, the center values of the mock data can be given as follows,
\begin{align}
    X_i^{\rm mock} = X^\text{fiducial}(t_i)+\delta X_i, \hspace{1cm}
    Y_i^{\rm mock} = Y^\text{fiducial}(t_i)+\delta Y_i. \label{5}
\end{align}
Our mock data consists of $N$ data points $\{X^\text{mock}_i, Y^\text{mock}_i, \sigma_{X,i}, \sigma_{Y,i} \}_{i=1}^N$, where each point includes astrometric positions and their associated uncertainties.

\begin{figure}
    \centering
    \begin{subfigure}[b]{0.49\textwidth}
        \centering
        \caption{Centroid from primary images}
        \includegraphics[width=\textwidth]{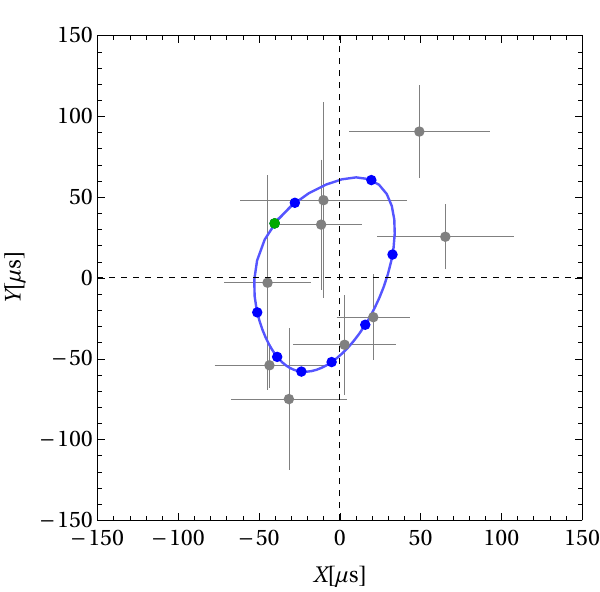}
        \label{}
    \end{subfigure}
    \hfill
    \begin{subfigure}[b]{0.49\textwidth}
        \centering
        \caption{Centroid from primary + secondary images}
        \includegraphics[width=\textwidth]{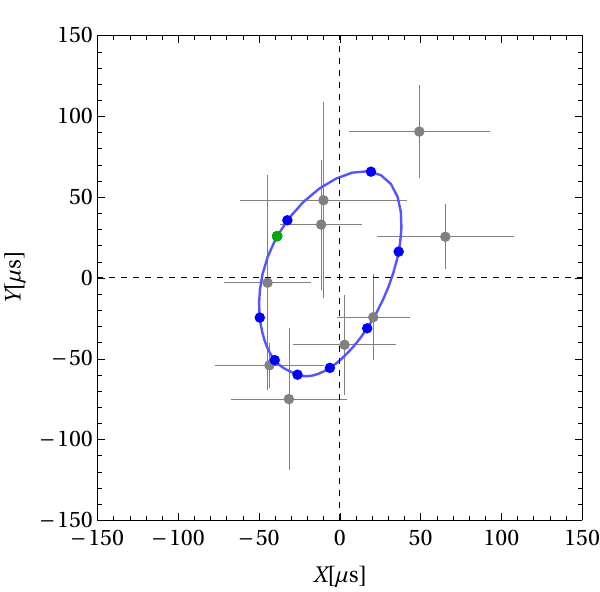}
        \label{}
    \end{subfigure}
    \caption{The best-fitted centroid tracks for the GRAVITY astrometric data, with the centroid only accounting for the primary images [panel (a)] and with the centroid accounting for both the primary and secondary images [panel (b)]. The astrometric data in gray was reported by GRAVITY collaboration. The blue curves are the best-fitted centroid tracks as parameters given in Tab.~\ref{best-fitting:fiducial values}.}
    \label{F3}
\end{figure}

\begin{figure}
    \centering
    \begin{subfigure}[b]{0.49\textwidth}
        \centering
        \caption{Mock data $D_0$ with $N=100$ and $f=0.4$}
        \includegraphics[width=\textwidth]{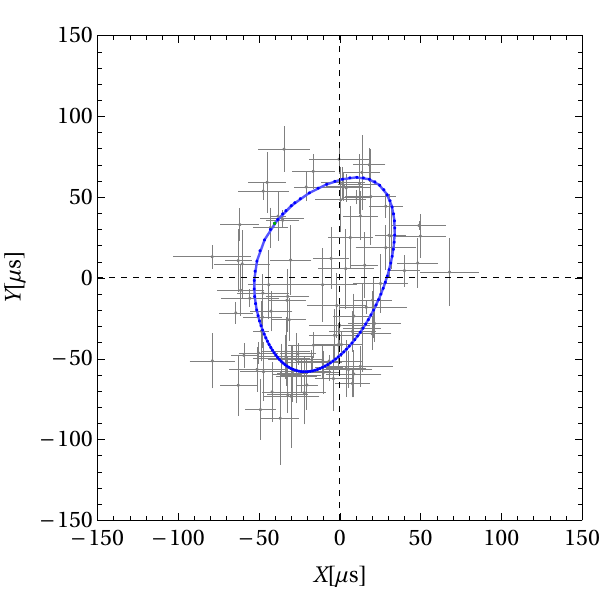}
        \label{}
    \end{subfigure}
    \hfill
    \begin{subfigure}[b]{0.49\textwidth}
        \centering
        \caption{Mock data $D_1$ with $N=100$ and $f=0.4$}
        \includegraphics[width=\textwidth]{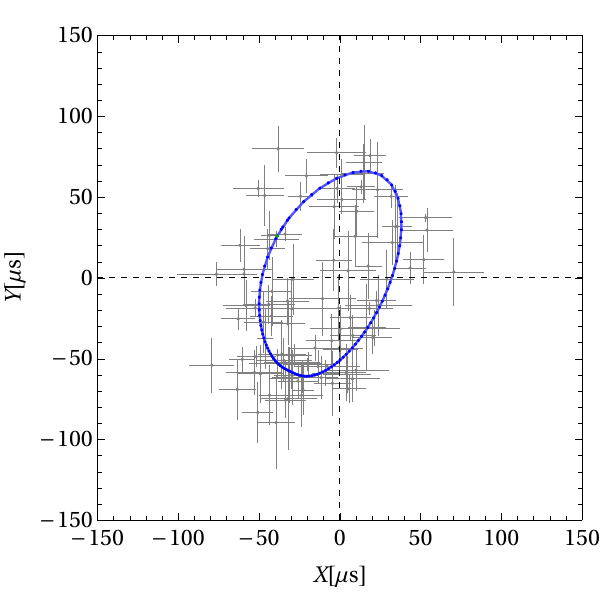}
        \label{}
    \end{subfigure}
    \caption{The mock data generated without [panel (a)] or with [panel (b)] contribution of secondary images for the centroid tracks. The gray points with error bar are the mock data. The blue curves represent the fiducial models with the parameters in Tab.~\ref{best-fitting:fiducial values}. In both panels, we use the same random seed.}
    \label{F4}
\end{figure}

In order to examine the detectability of secondary image, we consider two types of fiducial models,  $(X^\text{fiducial}(t), Y^\text{fiducial}(t))$: (i) the centroid tracks only accounting for primary image and (ii) the centroid tracks accounting for both primary and secondary images. The best-fit orbits are presented as blue curves in Fig.~\ref{F3} with the corresponding orbital parameters summarized in Tab.~\ref{best-fitting:fiducial values}. These two fiducial models are substituted in Eq.~(\ref{5}) to generate our mock data, denoted $D_0$ and $D_1$.
Specifically, the mock data $D_0$ is generated using only the primary image contribution to the centroid position, and the mock data $D_1$ includes both the primary and secondary images. 
\begin{table}[]
    \centering
    \caption{Best-fitting hotspot orbital parameters adopted as fiducial values}
    \setlength{\tabcolsep}{12pt}
    \renewcommand{\arraystretch}{1.3}
    \begin{tabular}{ccccccc}
    \toprule
         Fiducial Trajectory  & Mock data &   $\theta _{\rm inc}$  &  $r_0/M$  &  $\phi_0$  &  PA  &  $\omega/ \omega_{\rm K}$\\
    \hline
         Primary image only  & $D_0$ &  1.08  &  11.27  &  2.90  &  1.18  &  1.61\\
         Primary + secondary images  & $D_1$ &  1.08  &  12.42  &  2.83  &  1.15  &  1.85\\
    \hline
    \end{tabular}
    \label{best-fitting:fiducial values}
\end{table}
Using the above procedure to generate the mock data, we obtain the astrometric data as shown in Fig.~\ref{F4}. 
To investigate the impact of improved observations, we will generate enhanced mock data by increasing the sample size ($N>9$\footref{footnote}) and reducing measurement uncertainties (parameterized by a scaling factor $0<f<1$).


\subsection{Parameter inference in Bayesian framework}\label{mock}

We perform MCMC sampling using the emcee ensemble sampler to explore the parameter space within a Bayesian framework. In this work, we assume a circular orbit and fix the black hole mass to $4.3 \times 10^6 M_{\odot}$, since it is known precisely by monitoring the S2 star  \cite{massofblackhole}. Here, the orbital parameters are given by
\begin{equation}
    \Theta=(\theta_\text{inc},r_0/M,\phi_0,\text{PA},\omega/\omega_{\rm K})~,\label{parameters}
\end{equation}
where $\theta_{\text{inc}}$ denotes the inclination of the orbital plane, PA is the position angle of the orbital plane projected onto the observer's sky, and $\omega/\omega_{\rm K}$ represents the ratio between the orbital angular velocity and the Keplerian angular velocity.

The procedure of generating mock data is described in Sec.~\ref{mock data generation}, denoted ($X^\text{mock}_i$, $Y^\text{mock}_i$), in two different types $D_0$ and $D_1$.
Here, the astrometric ($X_i$, $Y_i$) in our mock data is assumed to be statistically independent as illustrated in Sec.~{\ref{mock data generation}}. Given the model parameters defined in Eq.~(\ref{parameters}) and our data, the log-likelihood function is written as
\begin{equation}
    \mathcal{L}(\Theta_\text{model})=-\frac{1}{2}\sum_{i=1}^N\left(\left(\frac{X_i^\text{mock}-X^\text{centroid}(t_i)}{\sigma_{X,i}}\right)^2+\left(\frac{Y_i^\text{mock}-Y^\text{centroid}(t_i)}{\sigma _{Y,i}}\right)^2\right)~, \label{like1}
\end{equation}
where $\bm X^\text{centroid}(t)$ is the centroid positions defined in Eq.~(\ref{3}). In this study, we consider two variants of the centroid positions: (i) the primary images dominant the radiation flux of the hotspot such that centroid position reduces to the tracks of the primary images, denoted $M_0$; (ii) the centroid positions are collectively determined by the primary and secondary images, and the higher-order image ($n>1$) is neglected, denoted $M_1$. By performing MCMC parameter inference and analyzing the statistical preference for a specific model ($M_0$ or $M_1$), we can explore the distinguishability of secondary images in the flare astrometric data. 
For MCMC sampling, we use 32 walkers, each evolved for $10^5$ steps. Uniform priors are adopted for all model parameters. 

In summary, the models used for MCMC parameter inference (Sec.~\ref{mock}) and generation procedure for the mock data (Sec.~\ref{mock data generation}) are summarized in Tab.~\ref{tab:def_mock_data}. 
\begin{table}[]
    \centering
    \caption{Definitions of the models and mock data used in this work.}
    \setlength{\tabcolsep}{12pt}
    \renewcommand{\arraystretch}{1.3}
    \begin{tabular}{ccp{11cm}}
    \toprule
    Symbol  &  Type  &  Description\\
    \hline
    $D_0$   &  Mock data  &  Simulated centroid data generated from trajectories only accounting for contributions from the primary images.\\
    
    $D_1$   &  Mock data  &  Simulated centroid data generated from trajectories accounting for contributions from both the primary and secondary images.\\
    
    $M_0$   &  Model   &  Centroid tracks constructed by the positions of the primary images:  $\bm X^\text{centroid}(t) = \bm X^{(0)}{(t)}$. \\ 
    
    $M_1$   &   Model  &  Centroid constructed by combining the flux-weighted positions of the primary and secondary images: $\bm X^\text{centroid}(t) = \bm X^{(0)}{(t)}\bar{F}^{(0)}_\nu+\bm X^{(1)}{(t)}\bar{F}^{(1)}_\nu$.\\
    \hline
    \end{tabular}
    \label{tab:def_mock_data}
\end{table}

\section{Results}\label{results}

To investigate whether secondary images can be detected in the future GRAVITY astrometric flare data, 
we use the mock data $D_0$ and $D_1$ as described in Tab.~\ref{tab:def_mock_data}, and consider each mock data containing $N=100$ astrometric data points, corresponding to the accumulation of the combined flare events reported by GRAVITY collaboration. 
This choice of sample size corresponds to an optimal sampling scale identified in the validation tests detailed in Sec.~\ref{Validation with mock data and choice of sample size}. 
Furthermore, to explore the impact of improved instrumental precision on the detectability of the secondary image, we will compare orbit fit for mock data with scaling factor $f=1$ (the current precision from GRAVITY datasets), and that for $f=0.4<1$ (the improved-precision case with uncertainties reduced to $40\%$ of the current values). 
Under these setups, we perform MCMC parameter inference for the mock data and centroid models, and analyze the model preference using the Bayesian Information Criterion (BIC). Here, the illustration of mock data and fitted models are summarized in Tab.~\ref{tab:def_mock_data}. 

\subsection{Validation with mock data and choice of sample size}\label{Validation with mock data and choice of sample size}

To validate the stability of our inference and determine an appropriate sample size for the mock data, we perform a series of tests with varying numbers of simulated data points $N$. For each choice of $N$, synthetic flare astrometric datasets are generated following the procedure described in Sec.~\ref{mock data generation}, and the same MCMC inference framework used in the main analysis is applied to each data. 

We tested several choices of the number of mock data points, $N=10$, $20$ ,$30$, ..., $130$, using the same MCMC procedure as in the main analysis. The results indicate that increasing $N$ improves the posterior constraints, but the gain becomes limited once $N$ reaches approximately $100$. This arises because while the sample size $N$ was increased, the total observation duration of each flare event was fixed at approximately one hour \cite{GRAVITY:2023data}, thereby restricting the trajectory to a single orbit. Hence, we adopt $N=100$ as the fiducial sample size in Sec.~\ref{results}. 

\subsection{Current observational precision ($f=1$)}

\begin{figure}
    \centering
    \begin{subfigure}[b]{0.49\textwidth}
        \centering
        \caption{Data without secondary images ($D_0$) with $f=1$}
        \includegraphics[width=\textwidth]{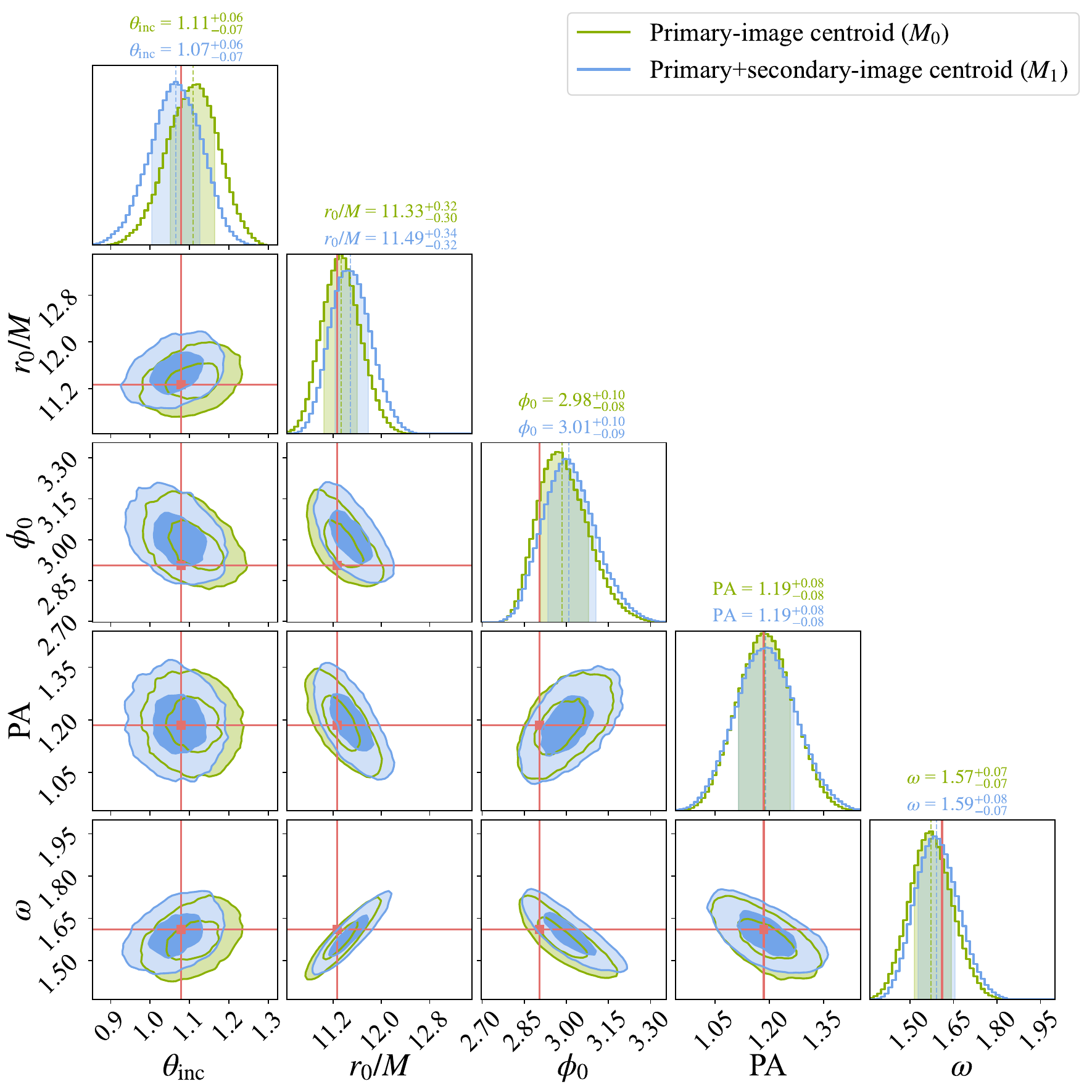}
        \label{}
    \end{subfigure}
    \hfill
    \begin{subfigure}[b]{0.49\textwidth}
        \centering
        \caption{Data with secondary images ($D_1$) with $f=1$}
        \includegraphics[width=\textwidth]{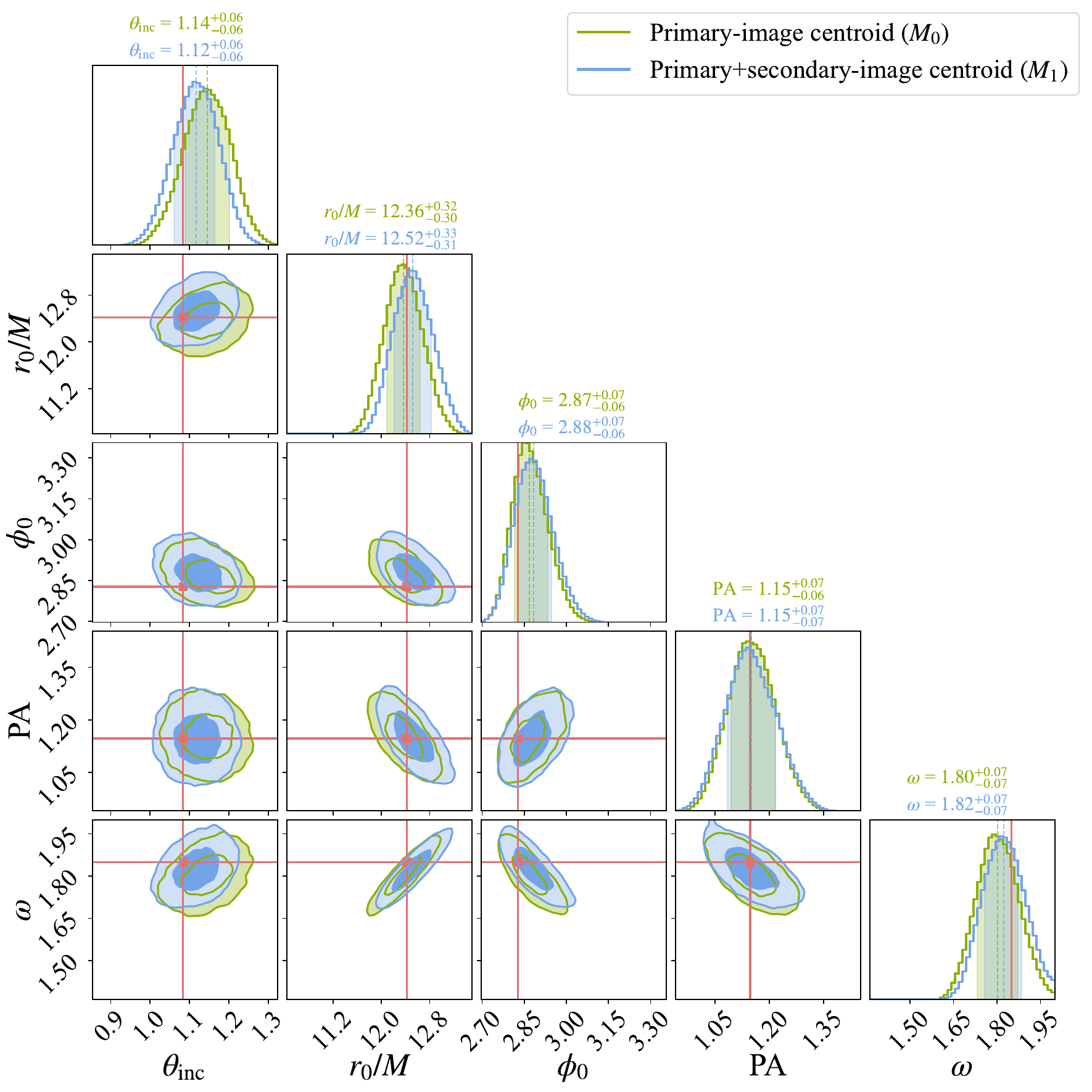}
        \label{}
    \end{subfigure}
    \caption{Corner plots of the posterior distributions for $f=1$. Panel (a) shows the results for the mock data $D_0$ without the contribution of secondary image, while panel (b) corresponds to the mock data $D_1$ including the contribution of secondary image. In each panel, the green and blue contours represent fits with $M_0$ and $M_1$, respectively. The red lines indicate the fiducial parameter values used to generate the corresponding mock data. 
    The mock data and fitted models are summarized in Tab.~\ref{tab:def_mock_data}. 
    }
    \label{fig:corner_f1}
\end{figure}

We first consider the scaling factor $f=1$, corresponding to the current observational precision of GRAVITY. 
We investigate whether the contribution of secondary images can be identified from the posterior constraints and whether the mock data $D_0$ and $D_1$ can be distinguished from their posterior distributions.

Fig.~\ref{fig:corner_f1} shows the posterior distributions of the model parameters for the mock data $D_0$ and $D_1$ fitted with the two centroid models $M_0$ and $M_1$. The illustration of mock data and fitted models are summarized in Tab.~\ref{tab:def_mock_data}. The resulting posterior median values and the corresponding uncertainties are summarized in Tab.~\ref{tab:f1}. 
For both $D_0$ and $D_1$, the fiducial parameter values are well recovered by both fitting models within the posterior uncertainties. 
The posterior distributions obtained under $M_0$ and $M_1$ show substantial overlap, indicating that the secondary image does not lead to a clearly distinguishable signature at the current precision. 
The correlations in the posterior distributions inferred for $D_0$ and $D_1$ are also qualitatively similar. 
Therefore, 
it is difficult to determine whether the mock astrometric data contain the signature of secondary images at the current GRAVITY astrometric precision.  





\begin{table}[]
    \centering
    \caption{Posterior distributions of the model parameters for the mock data for $f = 1$.}
    \setlength{\tabcolsep}{12pt}
    \renewcommand{\arraystretch}{1.3}
    \begin{tabular}{cc|ccccc}
    \toprule
    data & Model  &  $\theta_\text{inc}$  &  $r_0/M$  &  $\phi_0$  &  PA  &  $\omega/\omega_{\rm K}$  \\
    \hline
    $D_0$&$M_0$   &  $1.11^{+0.06}_{-0.07}$  &  $11.33^{+0.32}_{-0.30}$  &  $2.98^{+0.10}_{-0.08}$  &  $1.19^{+0.08}_{-0.08}$  &  $1.57^{+0.07}_{-0.07}$  \\
    
    $D_0$&$M_1$   &  $1.07^{+0.06}_{-0.07}$  &  $11.49^{+0.34}_{-0.32}$  &  $3.01^{+0.10}_{-0.09}$  &  $1.19^{+0.08}_{-0.08}$  &  $1.59^{+0.08}_{-0.07}$  \\
    
    $D_1$&$M_0$   &  $1.14^{+0.06}_{-0.06}$  &  $12.36^{+0.32}_{-0.30}$  &  $2.87^{+0.07}_{-0.06}$  &  $1.15^{+0.07}_{-0.06}$  &  $1.80^{+0.07}_{-0.07}$\\
    
    $D_1$&$M_1$   &   $1.12^{+0.06}_{-0.06}$  &  $12.52^{+0.33}_{-0.31}$  &  $2.88^{+0.07}_{-0.06}$  &  $1.15^{+0.07}_{-0.07}$  &  $1.82^{+0.07}_{-0.07}$  \\
    \hline
    \end{tabular}
    \label{tab:f1}
\end{table}

\subsection{Improved observational precision ($f=0.4$) \label{IVB}}

\begin{figure}
    \centering
    \begin{subfigure}[b]{0.49\textwidth}
        \centering
        \caption{Data without secondary images ($D_0$) with $f=0.4$}
        \includegraphics[width=\textwidth]{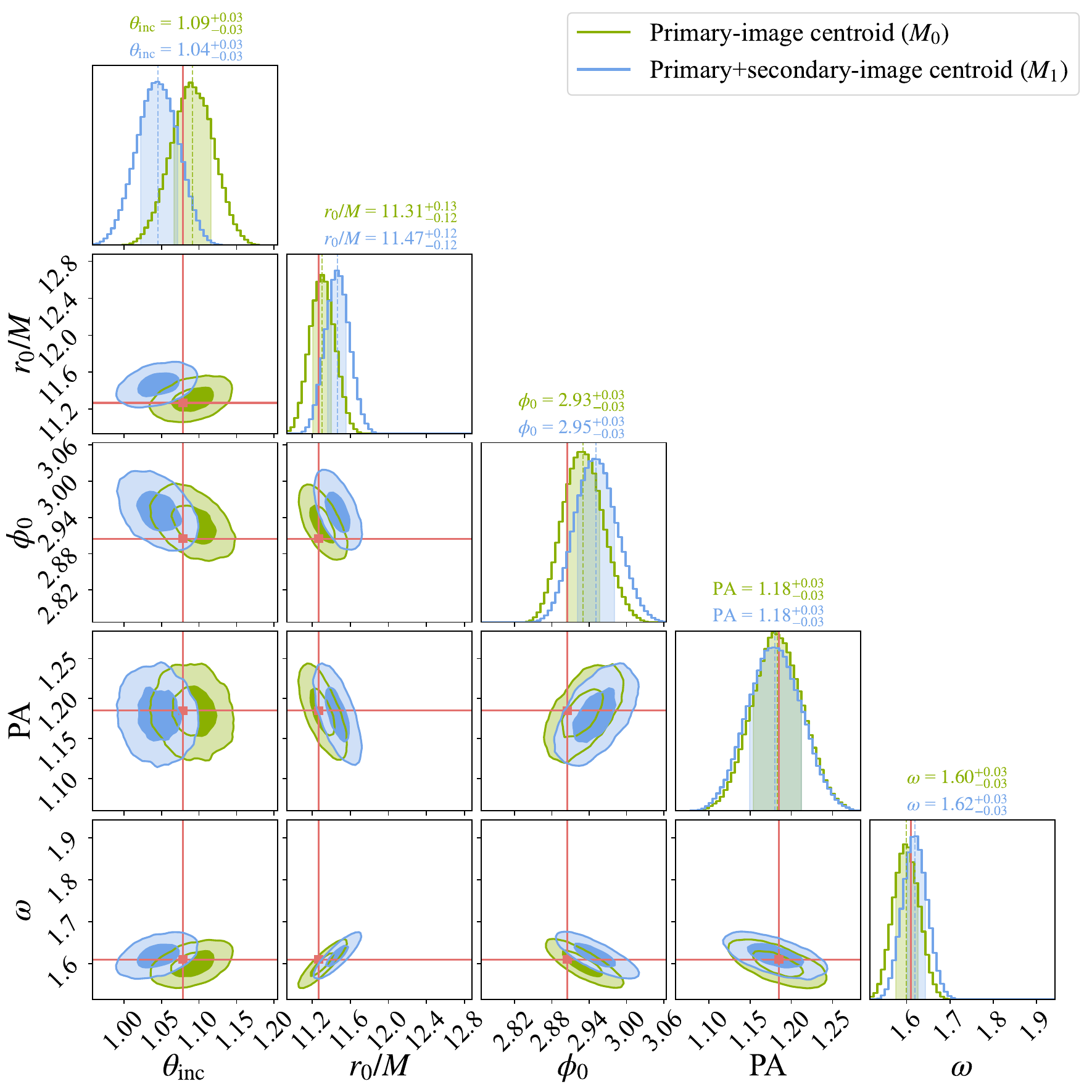}
        \label{}
    \end{subfigure}
    \hfill
    \begin{subfigure}[b]{0.49\textwidth}
        \centering
        \caption{Data with secondary images ($D_0$) with $f=0.4$}
        \includegraphics[width=\textwidth]{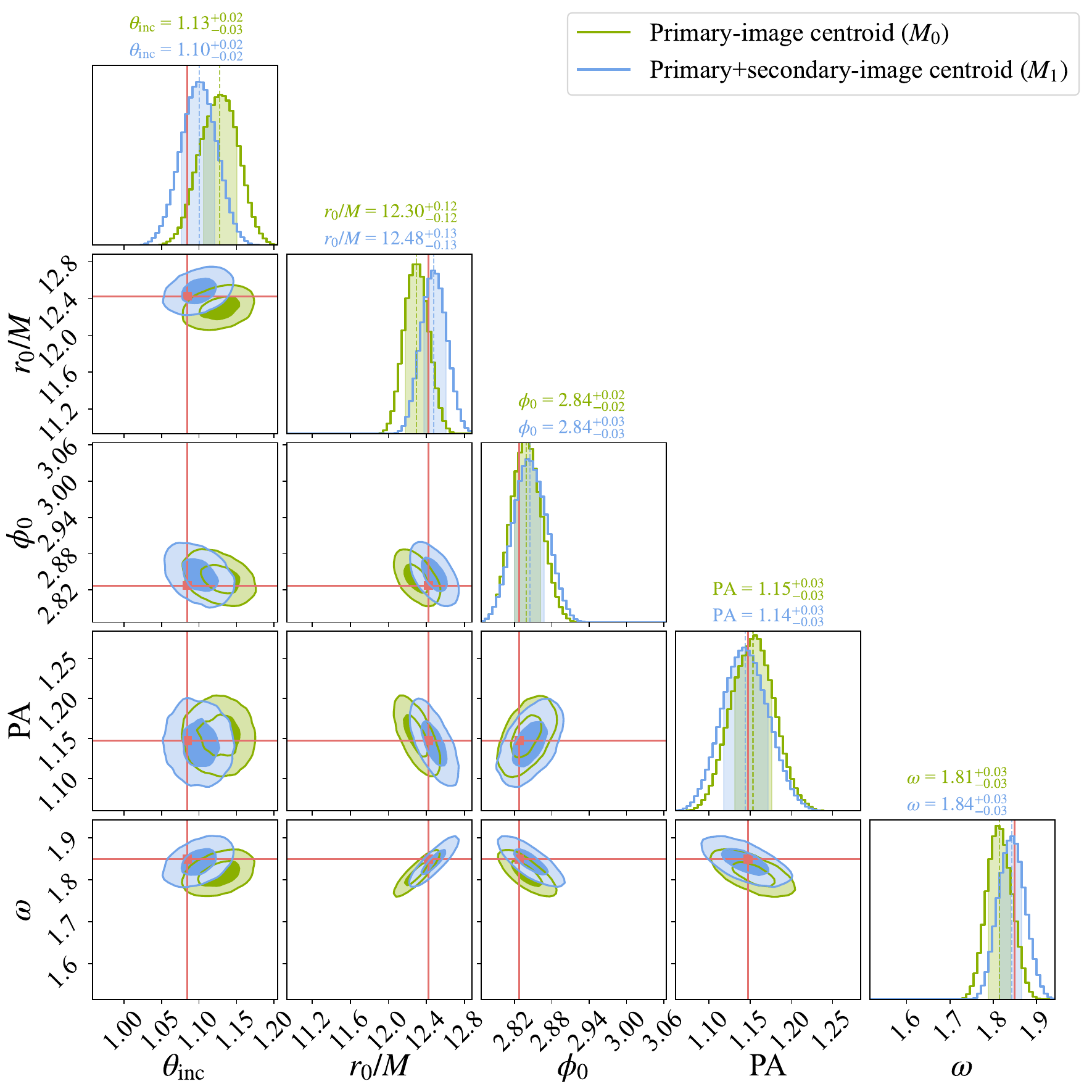}
        \label{}
    \end{subfigure}
    \caption{Corner plots of the  posterior distributions for $f=0.4$. Panel (a) shows results for the mock data $D_0$ without the contribution of secondary image, while panel (b) corresponds to the mock data $D_1$ including the contribution of secondary image. In each panel, green and blue contours represent fits with $M_0$ and $M_1$, respectively. The red lines indicate the fiducial parameter values used to generate the corresponding mock data. 
    The mock data and fitted models are summarized in Tab.~\ref{tab:def_mock_data}. 
    }
    \label{fig:corner_f04}
\end{figure}

We then consider the improved-precision case $f=0.4$, where the astrometric uncertainties are reduced to $40\%$ of the current GRAVITY astrometric precision. 
This case allows us to examine whether improved observational precision can enhance the distinguishability of the contribution of secondary image. 

Fig.~\ref{fig:corner_f04} shows the posterior distributions of the model parameters for the mock data $D_0$ and $D_1$ fitted with the two centroid models $M_0$ and $M_1$. The posterior median values and the corresponding uncertainties are summarized in Tab.~\ref{tab:f04}. 
Compared with the $f=1$ case, the posterior distributions are more concentrated and become more separated between different models $M_0$ and $M_1$, as expected from the reduced observational uncertainties. 
For both $D_0$ and $D_1$, the fiducial parameter values are recovered within the posterior uncertainties by both fitting models. 

The correlation in the posterior distribution still does not exhibit a characteristic signature, indicating that the secondary images could not produce a clearly distinguishable signature. 
The posterior distributions inferred from $D_0$ and $D_1$ also show similar overlap structures. 
While there are systematic offsets in several parameters, they are not sufficient to demonstrate a clear distinction between the two datasets. 
In this sense, a quantitative approach should be adopted to evaluate whether the mock astrometric data contain the contribution of secondary image.


\begin{table}[]
    \centering
    \caption{Posterior distributions for $f = 0.4$.}
    \setlength{\tabcolsep}{12pt}
    \renewcommand{\arraystretch}{1.3}
    \begin{tabular}{cc|ccccc}
    \toprule
    data & Model  &  $\theta_\text{inc}$  &  $r_0/M$  &  $\phi_0$  &  PA  &  $\omega/\omega_{\rm K}$  \\
    \hline
    $D_0$&$M_0$   &  $1.09^{+0.03}_{-0.03}$  &  $11.31^{+0.13}_{-0.12}$  &  $2.93^{+0.03}_{-0.03}$  &  $1.18^{+0.03}_{-0.03}$  &  $1.60^{+0.03}_{-0.03}$  \\
    
    $D_0$&$M_1$   &  $1.04^{+0.03}_{-0.03}$  &  $11.47^{+0.12}_{-0.12}$  &  $2.95^{+0.03}_{-0.03}$  &  $1.18^{+0.03}_{-0.03}$  &  $1.62^{+0.03}_{-0.03}$  \\
    
    $D_1$&$M_0$   &  $1.13^{+0.02}_{-0.03}$  &  $12.30^{+0.12}_{-0.12}$  &  $2.84^{+0.02}_{-0.02}$  &  $1.15^{+0.03}_{-0.03}$  &  $1.81^{+0.03}_{-0.03}$\\
    
    $D_1$&$M_1$   &   $1.10^{+0.02}_{-0.02}$  &  $12.48^{+0.13}_{-0.13}$  &  $2.84^{+0.03}_{-0.03}$  &  $1.14^{+0.03}_{-0.03}$  &  $1.84^{+0.03}_{-0.03}$  \\
    \hline
    \end{tabular}
    \label{tab:f04}
\end{table}

\subsection{Model comparison with Bayes factor}\label{Model comparison with BIC}

To quantify the statistical preference between the centroid models, we employ the BIC to evaluate the fits for the mock data ($D_0$ and $D_1$) and the models ($M_0$ and $M_1$). The illustration of mock data and fitted models are summarized in Tab.~\ref{tab:def_mock_data}. Here, the BIC is defined as
\begin{equation}
{\rm BIC}=k\ln(2N)-2 \mathcal{L}_{\rm max},
\end{equation}
where $k$ is the number of model parameters, $N$ is the sample size, and $2N$ is the total number of independent data components from the $X_i$ and $Y_i$,  and $\mathcal{L}_{\rm max}$ is the maximum log-likelihood. Since the difference in BIC values provides an asymptotic approximation to Bayes factor, we thus use $\Delta$BIC for model comparison, namely,
\begin{equation}
\Delta {\rm BIC} \equiv {\rm BIC}_{M_1}-{\rm BIC}_{M_0}.
\end{equation}
With this convention, a positive value favors the model $M_0$, while a negative value favors the model $M_1$.
The results are summarized in Tab.~\ref{tab:bic}. 
\begin{table}[]
    \centering
    \caption{Model comparison based on the BIC}
    \setlength{\tabcolsep}{12pt}
    \renewcommand{\arraystretch}{1.3}
    \begin{tabular}{cccccc}
    \toprule
       data & $f$ & $\mathrm{BIC}_{M_0}$ & $\mathrm{BIC}_{M_1}$ & $\Delta \rm BIC$ & Interpretation\\
       \hline
       $D_0$  &  1 &  249.730  &  252.581  &  2.851  &  Weakly favors $M_0$\\      
       $D_1$  &  1 &  250.944  &  251.504  &  0.560  &  No significant  preference\\
       $D_0$  &  0.4  &  249.798  &  261.579  &  11.781  &  Strongly favors $M_0$\\
       $D_1$  &  0.4  &  259.344  &  251.396  &  $-7.948$  &  Strongly favors $M_1$\\
       \hline
       \end{tabular}
    \label{tab:bic}
\end{table}

For the current precision $f=1$, even when increasing the sample size to $N=100$, the $\Delta\text{BIC}$ indicates only weak or negligible model preference. 
Specifically, for data $D_0$, we obtain $\Delta{\rm BIC}=2.851$, which weakly favors model $M_0$, consistent with the fact that $D_0$ is generated without the contribution of secondary image. 
And for data $D_1$, $\Delta{\rm BIC}=0.560$, indicating no significant preference between models $M_0$ and $M_1$. 
This suggests that, at the current GRAVITY astrometric precision, the BIC comparison does not provide significant evidence for identifying the contribution of secondary image.
When the astrometric uncertainties are reduced to $f=0.4$, the model preference becomes more pronounced. 
For data $D_0$, $\Delta{\rm BIC}=11.781$, strongly favoring model $M_0$, whereas for data $D_1$, $\Delta{\rm BIC}=-7.948$, strongly favoring model $M_1$. 
This result indicates that improved astrometric precision can enhance the statistical separability between the centroid models $M_0$ and $M_1$.

Based on $\Delta\text{BIC}$ for model comparison, we demonstrate that detecting the secondary images become feasible with the astrometric data of sample size $N=100$ and uncertainties reduced to 40\% relative to current capabilities.

\section{ Conclusions and Discussions}\label{Discussion and Conclusions}

In this work, we explore the detectability of the secondary images through the astrometric motion of flares near Sgr~A*,  using the mock data simulating future GRAVITY observations. The illustration of mock data and fitted models are summarized in Tab.~\ref{tab:def_mock_data}.
Specifically, we considered two contrasting mock data: $D_0$, generated from a centroid track accounting for the primary images, and $D_1$, generated from a centroid track including both the primary and secondary images. Namely, these two mock data simulate the situations in which future GRAVITY observations either exclude or include contributions from secondary images to astrometric motion of flares. 
We fit the mock data using two astrometric centroid models. Model $M_0$ assumes the hotspot flux is dominated by primary images, yielding a centroid track that coincides with the track of the primary images. Model $M_1$ incorporates contributions from both primary and secondary images, with the centroid computed as their flux-weighted image position.
Through model comparison with $\Delta\text{BIC}$, detection of the secondary images
becomes feasible with the astrometric data of sample size $N = 100$ and uncertainties reduced to
40\% relative to current capabilities.

Our phenomenological framework focuses on quantifying detectability thresholds rather than addressing the fundamental physical mechanism that could suppress or eliminate the secondary image signal. Exploring this specific mechanism is also of physical interests. The most straightforward explanation for the centroid tracks of flares being dominated by primary images is that the environment around Sgr~A* remains optically thick. Consequently, secondary images with longer light paths are more easily attenuated, resulting in $F^{(0)}_\nu \gg F^{(n)}_\nu$ for $n\geq1$. Besides, this may also provide insight into compact objects without a photon sphere \cite{Pelle:2024eyt,Igata:2025xxb}. 

Although the signature of secondary images, arising from  gravitational lensing for source situated in strong-field regime, can be physically distinguished from direct images dominated by accretion disk emission, they still provide limited information on the photon sphere of the black hole \cite{johnson2024black}. In this context, detecting secondary images serves as an essential first step toward probing higher-order images in the future, which will reveal more detailed features of the black hole spacetime geometry.

Our results suggest that the currently available GRAVITY flare astrometric data are not sufficient to unambiguously identify the contribution of secondary images. However, with more flare observations and further improvements in astrometric precision, it may become possible to detect secondary images through centroid tracks and model comparison. 
Future work could incorporate more realistic hotspot emission structures, gravity environment, and black hole geometries to assess the detectability of higher-order images in GRAVITY flare observations more systematically. 

\smallskip
{\it Acknowledgments}:  This work has been supported by the National Natural
Science Fund of China Grants (No. 12305073, No. 12347101, and No. 12275034). The authors thank Prof. Hai-Nan Lin for useful discussions.

\appendix

\bibliography{ref}

\end{document}